\documentclass[
               ]{jacow}

\usepackage[english]{babel}

\begin{document}

\title{CAPABILITIES AND LIMITATIONS OF NON-REDUNDANT APERTURE INTERFEROMETRY FOR BEAM SIZE MEASUREMENTS}

\author{U.~Iriso, L.~Torino\thanks{ltorino@cells.es}, A.~Turull Barot\thanks{alba2.turull@gmail.com}, ALBA Synchrotron (Spain), Cerdanyola del Vallès, Spain\\ 
B.~Nikolic, Cavendish Laboratory, Cambridge, United Kingdom\\ 
N.~Thyagarajan, Commonwealth Scientific and Industrial Research Organisation, Canberra, Australia\\ 
C.~Carilli, National Radio Astronomy Observatory, Socorro, NM, United States}

\maketitle

\begin{abstract}
Non-Redundant Aperture Interferometry (NRAI) is a beam characterization technique developed at ALBA in collaboration with radio-astronomy institutes. It enables the single-acquisition measurement of the full 2D transverse profile of the electron beam using visible synchrotron radiation. To better understand the technique limitations and performance, we performed extensive SRW simulations and compare them with experimental data. This paper presents the results of these studies, which define the capabilities and limits of NRAI applied to the current ALBA machine.


\end{abstract}

\section{INTRODUCTION}
The characterization of transverse beam dimensions is a key aspect in accelerators, as it provides a measurement of beam emittance. These measurements are usually performed using synchrotron radiation at light sources. The most common techniques used are either the X-ray pinhole \cite{Elleaume1995} or synchrotron radiation interferometry (SRI) \cite{Mitsuhashi1997}, if visible light is preferred. In the latter case, the measurement is typically unidimensional (two-aperture interferometry) and does not allow for full beam shape reconstruction in a single shot \cite{Torino2016}.

By using Non-Redundant Aperture Interferometry (NRAI), a technique widely used in astronomy, we have developed a method to acquire all transverse beam dimensions in real time with only one acquisition using visible light \cite{Nikolic2024, Torino:IBIC25-TUPMO09}.

The main idea is to acquire the interference pattern and analyze it in the Fourier domain where, due to the non-redundancy, each pair of pinholes interferes at a specific spatial frequency, allowing us to see distinct peaks. By calculating the visibility or the correlation—summing the intensity within these peaks and identifying the combinations that cause each peak while using the illumination as a free parameter—we can reconstruct the beam profile with a 2D Gaussian and obtain the beam parameters to convert them into laboratory coordinates.

This report describes the results of the NRAI method through Synchrotron Radiation Workshop (SRW) simulations to model the interferometer and the synchrotron radiation in a controlled environment, followed by the experimental validation at the Xanadu beamline \cite{SRW_EPAC98}.

\section{EXPERIMENTAL SETUP}
\begin{figure}
    \centering
    \includegraphics[width=1\linewidth]{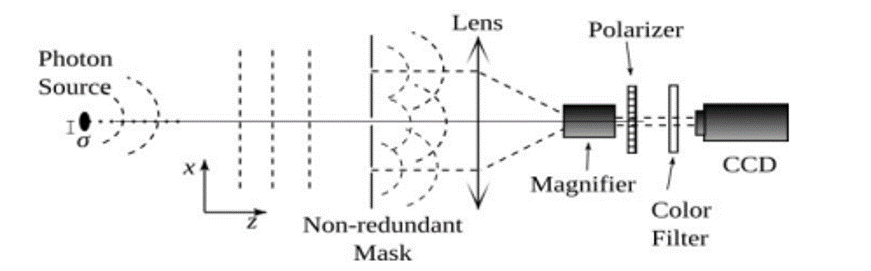}
    \caption{Experimental setup.}
    \label{fig:Syst}
\end{figure}

The experimental setup (Fig. \ref{fig:Syst}) selects only the visible light of the synchrotron radiation using a half-mirror located upwards with respect to the beam axis. The light then propagates through the Non-Redundant Mask (NRM). 

\begin{figure}[!h]
    \centering
    \includegraphics[width=.7\linewidth]{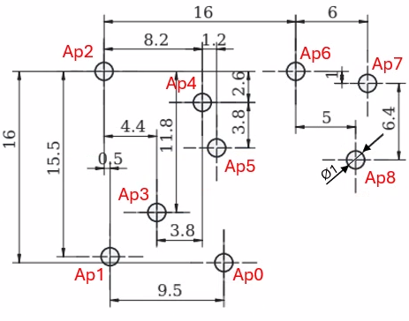}
    \caption{Schematic of the 9-hole NRA mask used at ALBA with the distances in mm.}

    \label{fig:Masks}
\end{figure}

Assuming the source is in the far-field limit, we can model the synchrotron radiation as plane waves passing through the NRM. These masks consist of a combination of pinholes—in our case, 5, 7, or 9—placed in a geometry where each baseline (the vector connecting two apertures) is unique. This ensures that each pinhole pair interferes at a unique spatial frequency in a given orientation, which reduces the risk of incoherence due to turbulence and non-uniformity in the radiation path. The 9-hole NRA mask used in this study is presented in Fig. \ref{fig:Masks}, while the 5 and 7-hole configurations are described in Ref.\cite{Torino:IBIC25-TUPMO09}. As shown in Fig.~\ref{fig:Syst}, the light then passes through a lens, a magnifier, a polarizer, and a color filter to select the desired polarization and wavelength, typically horizontal polarization and 400$\,$nm. Finally, the interference pattern is captured by a CCD camera.


\section{RESULTS}
In this section, the capabilities and limitations of this technique are discussed. To evaluate the performance of the NRAI, simulations were performed with SRW to establish resolution limits and the consistency of the procedure in ideal situations. The method was then validated experimentally at the Xanadu beamline to identify the limitations that appear in a real experimental environment.

\subsection{Simulation}
We simulated the Xanadu beamline at the ALBA Synchrotron using SRW. We model the source as a 2D Gaussian electron beam made up by 500 particles. We then generate the synchrotron radiation produced by each electron with a given starting position and propagate it to the camera. We sum the interferograms obtained from each electron to get the final result, and select the horizontally polarized monochromatic light (400$\,$nm). This is possible since the radiation produced by each electron is incoherent with respect to the others. 

\begin{figure}[h]
    \centering
    \includegraphics[width=1\linewidth]{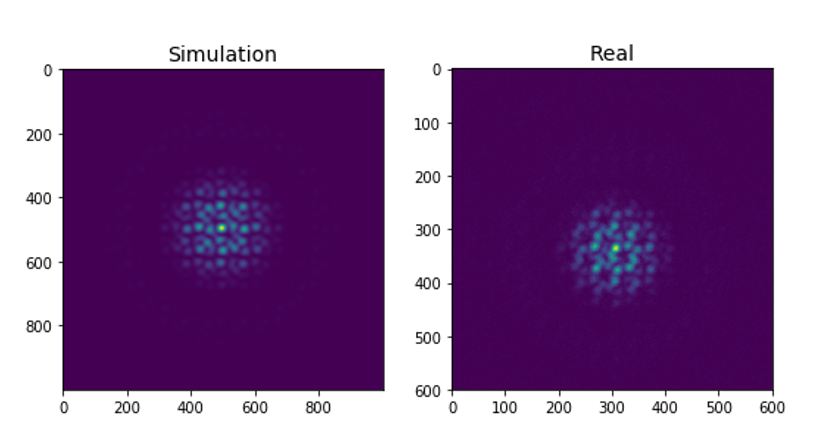}
    \caption{Raw interference images of the 5-hole NRA for the system simulated with SRW (left) and the experimental setup (right).}
    \label{fig:Diff}
\end{figure}

These simulations allow us to study the technique in a controlled environment without noise or instabilities. An example of the interferogram obtained by this simulation compared to the one captured in the real system is shown in Fig. \ref{fig:Diff} for a 5-hole NRM, demonstrating that the system produces the expected diffraction pattern, very similar to the experimental one. The slight discrepancy observed may be attributed to a difference in the mask's position between the simulation and the experimental setup, as well as effects from the pixel digitization and focus inaccuracies. 

\begin{figure}[h]
    \centering
    \includegraphics[width=1\linewidth]{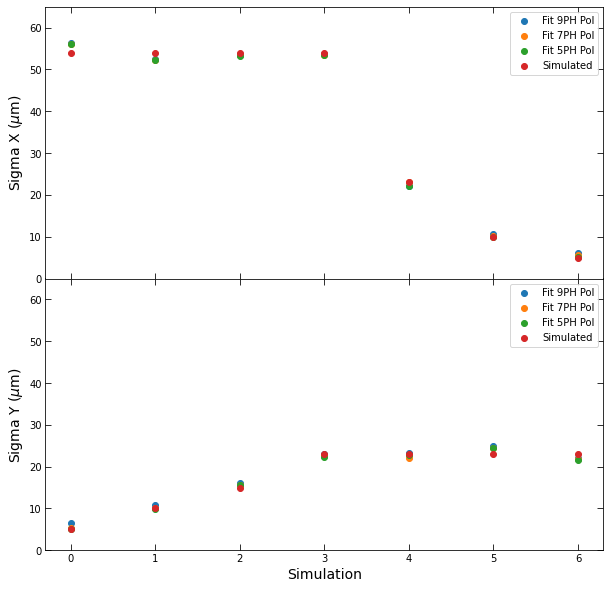}
    \caption{Simulated beam size scan results in the horizontal (upper) and vertical (lower) directions for NRA masks of 5, 7, and 9 holes.}
    \label{fig:BeamSizeScan}
\end{figure}

Using simulated interferograms allow us to improve the fitting algorithms to obtain the beam size results. We validate the accuracy of the fitting by performing a beam size scan. In this noise-free environment, we demonstrate that the algorithm can reconstruct very small beam sizes, such as 5$\,\mu$m, with low errors across all NRA masks used. As seen in Fig. \ref{fig:BeamSizeScan}, the results show that we can reconstruct small sizes in both horizontal and vertical dimensions. 

\begin{figure}[h]
    \centering    
    \includegraphics[width=1\linewidth]{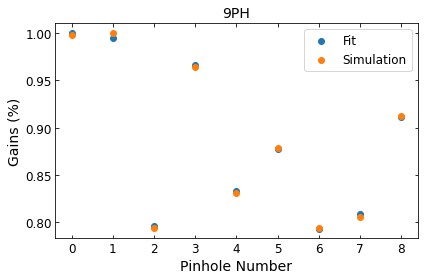}
    \caption{Reconstruction of the simulated gains with the values obtained by the fitting procedure for the 9-hole  mask.}
    \label{fig:SimG9}
\end{figure}

To further verify the robustness of the fitting results, we reconstructed the illumination profile (gains) using the simulated data. The results in Fig. \ref{fig:SimG9} show that we replicate the profile of the simulated data with remarkable precision, with a maximum relative error of only 0.5\% for the 9-hole mask.

The successful reconstruction of both beam dimensions and illumination gains confirms the robustness of the NRAI algorithm.

\subsection{Experimental Results}
We perform a vertical beam size scan to verify the ability to measure beam sizes using the 5, 7 and 9-holes masks.
Using the skew magnets, we reduce the coupling to the minimum and reach a beam size of around 20$\mu$m. Then, we apply a coherent vertical beam excitation using the bunch-by-bunch feedback system to increase until roughly 40$\mu$m. 

\begin{figure}[h]
    \centering    
    \includegraphics[width=1\linewidth]{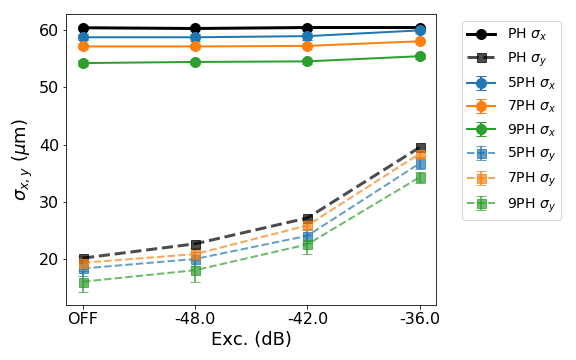}
    \caption{Horizontal (upper) and vertical (lower) dimensions of the reconstructed beam, for 5, 7, and 9-holes masks and the pinhole while changing the gain of  the vertical excitation.}
    \label{fig:XanaduXY}
\end{figure}

Figure~\ref{fig:XanaduXY} shows the measurements taken with each mask and compares them with the reference increments obtained using the FE21 x-ray pinhole~\cite{Iriso:IBIC22}. 
The results are consistent for all the masks. A systematic offset is observed for the 9-hole mask, which may be attributed to inaccuracies in the physical pinhole distances. All beam size results were calculated using theoretical model values. We note that the x-ray pinhole is in another location in the ring and the beamsizes should not be the same.



\begin{figure}[h]
    \centering    
    \includegraphics[width=1\linewidth]{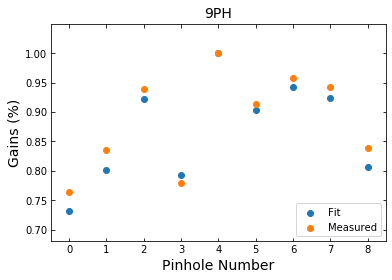}
    \caption{Reconstruction of the measured gains in the laboratory with the values obtained by the fitting procedure for the 9-hole NRA mask.}
    \label{fig:RealGains9}
\end{figure}

To further validate our results, we also performed gain reconstruction on the experimental data. Figure \ref{fig:RealGains9} compares the reconstructed gains for the 9-hole mask with the direct measurements, showing a maximum error of $3.5\%$. This confirms that the technique effectively compensates for synchrotron radiation inhomogeneity, even under non-ideal experimental conditions.

 
\begin{figure}[h]
    \centering    
    \includegraphics[width=1\linewidth]{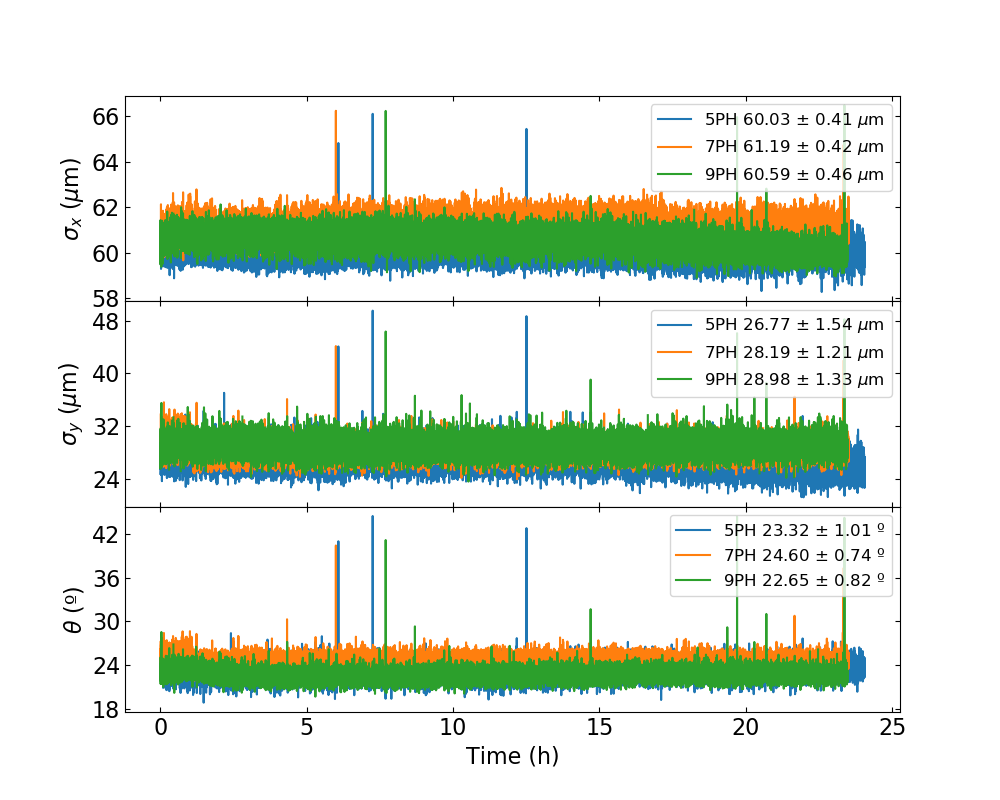}
    \caption{Beam transverse dimensions acquired in real time during 24h. Horizontal size (upper), vertical size (middle), and angle (lower). A blow up of the beam at around 5h of less than $10\%$ is observed in vertical size and angle (see zoom in Fig.~\ref{fig:bump}).}
    \label{fig:LongAcq}
\end{figure}

\begin{figure}[h]
    \centering    
    \includegraphics[width=1\linewidth]{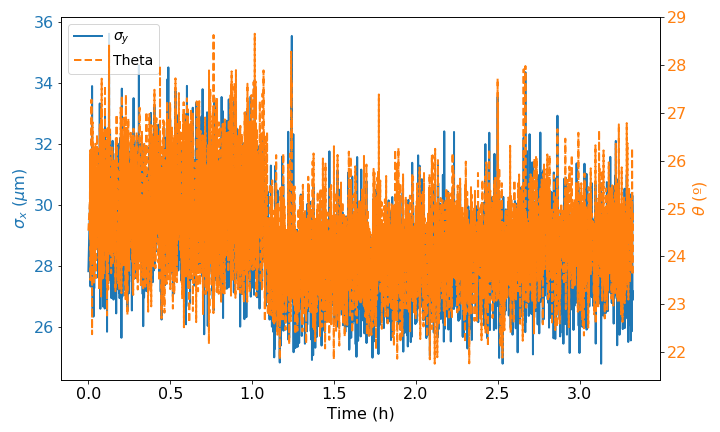}
    \caption{Increasing in vertical beam size and angle measured with the 7 NRM.}
    \label{fig:bump}
\end{figure}

For day-to-day operations, we created a tool to obtain beam dimensions in real time. This allows to track the long-term evolution and proves its availability as a routine tool for operation. Figure~\ref{fig:LongAcq} shows an acquisition of 24h with three different masks. Data were taken on different days, but always during  user operation conditions. The agreement within the three masks is very good. Note that the largest rms oscillations is less than 5\%, and it corresponds to the vertical (i.e. smallest) beam size. 
A small bump (beam size increase of $\sim$10\%) is visible for the vertical dimension and the angle, at the beginning of the acquisition with the 7 pinholes mask,  a zoom is presented in Fig. \ref{fig:bump}. This bump was due to noise problems in our Bunch By Bunch feedback system, which produced a small beam excitation. 

\section{CONCLUSION}
This report shows the extensive tests performed to analyze the NRAI technique applied to beam characterization for reconstructing transverse beam dimensions. The results of the SRW simulations show that the method is robust enough for application to the real diffraction system at the ALBA Synchrotron. We evaluated the functionality of the method through a beam size scan and the reconstruction of the gains, which demonstrated a high capability to capture the overall beam tendencies.

Finally, we are now able to perform real-time measurements of the beam dimensions using this technique over several days, providing an additional tool for beam characterization. This represents a very promising technique for the future ALBA II upgrade, and we will optimize our visible light diagnostics beamline accordingly to fully exploit its potential.


\begin{thebibliography}{9}

\bibitem{Elleaume1995}
    P.~Elleaume, C.~Fortgang, C.~Penel, and E.~Tarazona, 
    ``Measuring Beam Sizes and Ultra-Small Electron Emittances Using an X-Ray Pinhole Camera'', 
    \textit{J. Synchrotron Radiat.}, vol. 2, no. 4, pp. 209--214, Jul. 1995. 

\bibitem{Mitsuhashi1997}
T. Mitsuhashi, 
\textit{Spatial Coherency of the Synchrotron Radiation at the Visible Light Region and its Application for Vertical Beam Profile Measurement}, 
APS Meeting Abstracts, p. 3 (1997).

\bibitem{Torino2016}
L. Torino and U. Iriso, 
\textit{Transverse beam profile reconstruction using synchrotron radiation interferometry}, 
Phys. Rev. Accel. Beams \textbf{19}, 122801 (2016).

\bibitem{Nikolic2024}
B. Nikolic, C. Carilli, U. Iriso, N. Thyagarajan, and L. Torino, 
\textit{Two-dimensional synchrotron beam characterization from a single interferogram}, 
Phys. Rev. Accel. Beams \textbf{27}, 112802 (2024).

\bibitem{Torino:IBIC25-TUPMO09}
   L. Torino, B. Nikolic, C. Carilli, N. Thyagarajan, and U. Iriso,
   \textquotedblleft{Exploiting Non Redundant Aperture Interferometry as a Diagnostics Tool for Synchrotron Light Characterization}\textquotedblright,
   in \emph{Proc. IBIC’25}, Liverpool, UK, Sep. 2025, pp. 475--479.
   \doi{10.18429/JACoW-IBIC2025-TUPMO09}   

\bibitem{SRW_EPAC98}
O. Chubar and P. Elleaume, ``Accurate And Efficient Computation Of Synchrotron Radiation In The Near Field Region,'' in \textit{Proc. of the 6th European Particle Accelerator Conference (EPAC98)}, Geneva, Switzerland, 1998, pp. 1177--1179.

\bibitem{Iriso:IBIC22}
    U.~Iriso, A.A.~Cazorla, I.~Mases, and A.A.~Nosych, 
    ``PSF Characterization of the ALBA X-Ray Pinholes'', 
    in \textit{Proc. IBIC'22}, Krakow, Poland, Sep. 2022, pp. 421--425. 
    

\end{thebibliography}
\end{document}